\documentstyle[11pt,newpasp,twoside]{article}
\def\edcomment#1{\iffalse\marginpar{\raggedright\sl#1\/}\else\relax\fi}
\reversemarginpar
\begin{document}
\title{Diffusion and Settling in Ap/Bp Stars}
 \author{Sylvain Turcotte}
\affil{Lawrence Livermore National Laboratory, L-413, P.O. Box 808, Livermore, CA, 94551, USA}

\begin{abstract}
Ap/Bp stars are magnetic chemically peculiar early A and late B type stars 
of the main sequence. They exhibit peculiar surface abundance anomalies
that are thought to be the result of gravitational settling and
radiative levitation. 
The physics of diffusion in these stars are 
reviewed briefly and some model predictions are discussed. 
While models reproduce some observations reasonably well, more work is
needed before the behavior of diffusing elements in a complex magnetic
field is fully understood.
\end{abstract}

\section{Introduction}

Ap/Bp stars are chemically peculiar stars in which in which relatively
strong magnetic fields have been measured (see 
Ryabchikova~1991; see also the many other papers
devoted to these stars in these proceedings).
They are early A and late B stars, ranging from 8000 to 15000~K in
effective temperature, in which large abundance anomalies have been
found. Their magnetic fields typically of the order of 1 to 10~kG.

Ap/Bp stars, while similar in many ways to other main sequence chemically peculiar stars,
have quite specific abundance anomalies.  They are typically enriched in
rare earth elements and some lighter elements such as silicon.  They have been found
to feature surface composition inhomogeneities (e.g. Kushnig et al.~1999, see also Hatzes~1996) 
and to have a stratified atmospheric composition (e.g. Ryabchikova et al.~2002). 
As the main difference between
them is the presence of magnetism in Ap stars, the understanding of the
formation and evolution of the chemical composition in Ap stars 
may yield a new understanding of the structure of the magnetic fields.

Some of the cooler Ap stars are pulsating, the so-called roAp stars. 
Those stars are within the boundaries of the classical instability strip
where pulsations are driven by the opacity of helium. As diffusion
leads to helium deficiencies, it is a challenge to ensure that the required
opacity is maintained in the driving region for 
the $\kappa$-mechanism to be efficient. Therefore, the modeling of
the stratification in Ap stars is of central importance to the study of
roAp stars.

\section{Diffusion in Stars}

Diffusion is defined here as the relative drift of elements with respects
to each other. This drift occurs when forces acting on the
the components of the stellar plasma differ from species to species.
The term settling refers to the inward drift of heavier species with respect to light
species under the action gravity only.
In this section I will briefly describe the main physical processes that
lead to (or inhibit) diffusion in stars, starting first with a general
description, followed by a more specific discussion of diffusion in the
presence of magnetic fields.

The types of stars in which diffusion occurs depend on how fast
diffusion can affect the abundances and what are the competing
processes. Stars with appreciable surface chemical peculiarities are
slowly rotating stars that feature low mass loss and shallow surface
convection zones, i.e. early F to late B main sequence stars (Michaud
et al.~1976), white dwarfs (Fontaine \& Michaud~1979), and hot horizontal branch stars 
(Michaud, Vauclair, \& Vauclair~1983).  Long lived low mass stars
can be significantly affected by diffusion in the core (Proffitt \&
Michaud~1991). 

\subsection{Diffusion in Non-magnetic Stars}

In non-magnetic and slowly rotating stars, spherical symmetry can be
assumed. Diffusion will lead to vertical (radial) stratification of the elements
but not to inhomogeneities on the surface. Diffusion in non-magnetic
stars has been studied extensively in the past (see Michaud et al.~1976; Richer, Michaud, \&
Turcotte~2000; and references therein) and so will be described here only very briefly.

The variation of composition for an element $i$ with time can be
calculated from the continuity equation
\begin{equation}
  {\partial c_i\over\partial t}= -{\partial\over\partial r} \left( c_i v_D(i) \right)    \, ,
\end{equation}
where the diffusion velocity is expressed as (Babel \& Michaud~1991a)
\begin{equation}
  v_D(i) =  D_i \left\{-{\partial c_i\over\partial r}+{A_im_p\over
kT}(g_{\rm rad; i}-g) + \alpha {\partial\ln T\over\partial r} + ... \right\} ,
\end{equation}
or a similar expression.
The diffusion coefficient $D_i$ is mainly determined by collisional cross
sections and, as the Coulombian interactions dominate, on the degree of ionization of species $i$.
It is an average of the diffusion coefficients of the individual
ionization states of species $i$. The term including the temperature
gradient is the thermal diffusion term. Its coefficient $\alpha$ is
also determined by collisional cross-sections.
One may add a turbulent diffusion coefficient to $D_i$ to account for
turbulent particle transport, such as due to convection.

The most important factors that determine the 
amplitude and sign of $v_D(i)$ is the net effect of the inward gravitational
acceleration and the outward radiation acceleration, $g_{\rm rad; i}-g$.
If $g-g_{\rm rad; i} \geq$0, then the species will levitate, whereas it will
sink if $g-g_{\rm rad; i \leq}$0. The radiative acceleration $g_{rad}$ is
highly dependent on the ionization state and the atomic properties of the
species. 
In the absence of mass loss, the stratification and the photospheric abundances are determined in the
most part by the combined effects of gravity, radiation pressure and
convection (or mixing).

The radiation acceleration on a species decreases when its abundances
increases because of saturation. It also depends to a lesser degree on the abundance
of other elements because of the competition for photons in overlapping lines for example
(Michaud et al.~1976; Richer et al.~1998).

Where the diffusion time scales are short, the abundances will reach an 
equilibrium (i.e. a null total velocity). It is an appropriate assumption in the atmosphere 
of Ap stars when evolutionary effects are neglected, i.e. the static
case.  In the simplest possible case, where only gravity and radiation pressure 
are taken into account, the equilibrium will be reached 
when the composition has evolved so that the radiative acceleration
balances gravity. The equilibrium solution is in fact more complicated
due to the contributions of other processes, including mass loss.

There are other microscopic processes that may have to be considered in the relatively cool and
outer regions of stars if they are stable. 
One is ambipolar diffusion through which protons and electrons drift with respect to hydrogen atoms which would in turn 
affect the drift of other species  (Babel \& Michaud~1991c). Another
that has so far not been shown to be significant in Ap stars is
light induced drift which can arise from an anisotropic radiation field in spectral lines   
(LeBlanc \& Michaud~1993). 

\subsection{The Effect of Magnetic Fields}

Magnetic fields have several consequences,
direct and indirect, on how diffusion will affect the chemical composition.

{\sl Effect on drift of ionized species.}\  The magnetic field reduces
the diffusion coefficients in the direction perpendicular to the
magnetic field lines (Michaud, M{\'e}gessier, \& Charland~1981). This effect is
proportional to the strength of the field and to the charge of the
species. Neutrals do not feel the field while ions do. As a result, 
the drift of ionized species across field lines may be impeded and they
might have a tendency to drift along the field lines. 

{\sl Effect on radiative accelerations.}\  Radiative accelerations are
mostly due to absorption in spectral lines. In the presence of a
magnetic field the degeneracy of the lines may be lifted. In lines that are
saturated, this will reduce the saturation and increase the radiative
acceleration by factors up to 50 (Borsenberger, Michaud, \& Praderie~1981; Alecian \&
Stift~2002). A smaller effect is that an horizontal component to radiative
acceleration is induced which can potentially have significant effects
on some species (Babel \& Michaud~1991b).

{\sl Effect on convection.}\  Convection can be suppressed in the
presence of strong vertical field lines.  Balmforth~et~al.~(2001) 
have shown that convection would be expected to be strongly suppressed
at the magnetic poles for stars with a field as small as $10^3$~G. At
the magnetic equator however, they argue that the magnetic restoring
force is not able to compete against buoyancy and that convection
is not suppressed, although the field might maintain some coherence in
fluid motions.  Turbulent motions that occur above the convection zone
in non magnetic stars may be stabilized in in the presence of the
magnetic field.

The combined effects of introducing magnetic fields in diffusion
calculations is that it breaks the spherical symmetry leading to surface
variations of abundances. This creates a map of the surface magnetic
field that can be decoded with a proper modeling of horizontal abundance
inhomogeneities with respect to the geometry of the magnetic field.
It will also lead to a different vertical stratification than in the
non-magnetic case because of the changes in mixing and radiative accelerations.

\section{Application to Ap/Bp Stars}

In this section we will review some results and expectations for Ap/Bp
stars. 

The goal is, naturally, for models of Ap/Bp stars to be able to reproduce  surface
inhomogeneities comparable to observations, in amplitude and in
morphology, as well as predicting the correct vertical stratification.
Because of the complexity of the problem, most models have included many
simplifications, e.g. treating the magnetic poles and equator,  where the
field lines are either vertical or horizontal, 
separately, or assuming a very simple magnetic geometry,
However simplified, these models provide predictions as to which
elements should be over- or under-abundant depending on specific field
geometries that can be compared to observations.

\subsection{Magnetic Poles - Vertical Field Lines}

As mentioned above, the convective
structure at the poles and equator is different. Convection is
expected to be suppressed, or at the very least substantially reduced,
at the poles. On the other hand, the vertical field lines do
not impede mass loss whereas the horizontal field lines at the equator
trap the wind but allow convection to occur.
At the poles, diffusion in Ap stars is comparable to diffusion
in hot non-magnetic stars, such as the HgMn stars.
The vertical stratification will mainly determined by the competition between
diffusion and mass loss with no impact from the field lines as they are
parallel to the motions. Radiative forces are affected by the magnetic
field as discussed above.

Babel \& Michaud~(1991a) produced a parameter free model for the star
53~Cam that showed that simple diffusion
is not sufficient to reproduce the observed surface abundance
inhomogeneities
and that another physical process, most probably mass loss, is needed.
They estimated that a mass loss of the order of $10^{-12}$-$10^{-14}$
M$\odot$\,yr$^{-1}$ is required. Mass loss of that order would be typical of A
stars in general, magnetic or not.
Babel~(1993) added mass loss to the model and 
claims to reproduce the observations qualitatively.
Vauclair, Dolez, \& Gough~(1991) examined the behavior of helium at the
poles of an Ap star to see if enough helium could be maintained in 
superficial regions to account for the pulsations of roAp stars (helium has a very small 
$g_{\rm rad}$ at normal abundances and therefore always settles). They
show that a helium reservoir can indeed be maintained with a mass loss
of the order of $10^{-13}$~M$\odot$\,yr$^{-1}$.

The composition of He depends only on the strength of the mass loss, but for other
elements a lot will depend on the profile of the radiative acceleration
for each species. Some species may be preferentially lost in the wind
while others will accumulate in the line forming regions. Species that 
aren't radiatively supported in the upper atmosphere will not be dragged
by a metallic wind in which H and He are not expelled (Babel~1994).

In such models the mass loss rate is essentially a free parameter since
the values needed to affect diffusion are below the threshold of
detectability. However, Ap stars offer the opportunity of constraining the
mass loss by comparing predicted and observed polar abundances, provided
the radiative accelerations are known with sufficient accuracy.

Recently, an effort has been initiated (Michaud et al.~2002) 
where consistent diffusion of
multiple species as has been done in non-magnetic stars for several
years (Richer et al.~2000, and references therein). Their
calculations do not take the effects of magnetism on diffusion apart
from a reduction in mixing relative to Am stars.

\subsection{Magnetic Equator - Horizontal Field Lines}

Wherever the field lines are nearly horizontal, convection is thought to occur 
but mass loss is not. The magnetic field should stabilize the atmosphere and 
allow stratification to form above the convection zone.
Ions and neutrals will behave differently as the
vertical motions cross field lines.

The magnetic fields in Ap stars are low enough so that their effect on diffusion 
is expected to be limited to the most superficial regions of the stars.
Horizontal fields stronger than $10^5$~G would be needed to prevent
species from being pushed upward in the line forming region 
(Michaud et al.~1981). 

As a result, elements that are not radiatively supported, such as He or O, will be
depleted from the equatorial regions. Elements that are pushed outward
by the radiative forces will likely be trapped by the field lines and
will accumulate and be observed as being overabundant. 

In the cases where there is a large fraction of neutrals and where the
neutral and ions do not behave similarly, i.e.
the neutral being supported and ions not and vice versa, the net effect can 
be more complicated.
For elements where the neutral is radiatively supported but the first
ionized state is not, the case of Si is discussed by Michaud et
al.~(1981), the neutral population would be pushed upward
across the field lines but the ionized population would settle. When it
does so its vertical drift will be impeded by the field lines. It would
tend to diffuse horizontally, and form a band around the region where
the field lines are horizontal. This effect would only be seen if the
reservoir for that element is limited, meaning that the element would
be radiatively supported only in the upper
fraction of the envelope. 

The detailed distribution of elements in equatorial or intermediate
regions (i.e. where the field is horizontal or oblique) will depend
heavily on the details of horizontal diffusion, which remains
uncertain.

\section{Conclusion}

Theoretical predictions for the vertical and horizontal stratification
in Ap/Bp stars are now starting to become more precise. However much
still lies ahead before the predictions can be considered reliable.
Some recent observations of abundance patterns in Ap stars can't be easily
reconciled with current theoretical expectations
(e.g. Strasser, Landstreet, \& Mathys~2001 for HD~187474). They present evidence of
polar overabundances of some elements (Si, Cr, Mn),
elements that are expected to supported by
radiative pressure and therefore should be lost to the wind. On the
other hand they find other abundances that are in good agreement with
theoretical expectations, e.g. oxygen depletion, enrichment in rare
earth elements. As they point out, the absence of detailed models
prevents meaningful comparisons with their observations.

Alecian \& Stift~(2002) show that one must be wary 
of any theoretical predictions based on incomplete models. 
Nonetheless, as more physics is incorporated in the models, and improved atomic data
lead to more accurate diffusion coefficients and radiative
accelerations, comparing the models to the observations will help us
understand how convection and mixing is affected by magnetism and how mass
loss takes place in A stars.

A major modeling effort needs to be done to couple the observations to
the underlying physical processes. taking into account the 3-D transport
processes and radiative transfer requires a large code development
effort which is however possible given today's computing power.

\acknowledgements
I wish to thank Georges Michaud for his helpful comments and suggestions
on a previous version of this paper.
This work was performed under the auspices of the U.S.
Department of Energy, National Nuclear Security Administration by the
University of California, Lawrence Livermore National Laboratory under
contract No.W-7405-Eng-48.

\references
\reference Alecian, G., \& Stift, M. J. 2002, \aap, 387, 271
\reference Babel, J. 1993, in Peculiar Versus Normal Phenomena in A-Type
and Related Stars, ASP Conf. Ser. 44, eds M. M. Dworetsky, F. Castelli,
\& R. Faraggiana (ASP:San Francisco), 458
\reference Babel, J. 1994, \aap, 301, 823
\reference Babel, J., \& Michaud, G. 1991a, \apj, 366, 560
\reference Babel, J., \& Michaud, G. 1991b, \aap, 241, 493
\reference Babel, J., \& Michaud, G. 1991c, \aap, 248, 135
\reference Balmforth, N. J., Cunha, M. S., Dolez, N., Gough, D. O., \& Vauclair, S. 2001, \mnras, 323, 362
\reference Borsenberger, J., Michaud, G., \& Praderie, F. 1981, \apj, 243, 533 
\reference Fontaine, G., \& Michaud, G. 1979, \apj, 231, 826
\reference Hatzes, A. P. 1996, in Stellar Surface Structure, IAU Symp. 176, eds K. G. Strassmeier \& J. L. Linsky (Kluwer:Dordrecht), 305
\reference Kushnig, R., Ryabchikova, T. A., Piskunov, N. E., Weiss, W. W., \& Gelbmann, M. J. 1999, \aap, 348, 924
\reference LeBlanc, F., \& Michaud, G. 1993, \apj, 408, 251
\reference Michaud, G., Charland, Y., Vauclair, S., \& Vauclair, G. 1976, \apj, 210, 447
\reference Michaud, G., M{\'e}gessier, C., \& Charland, Y. 1981, \aap, 103, 244
\reference Michaud, G., Brassard, P., Richer, J., Richard, O., \&
Fontaine, G. 2002, in Radial and Nonradial Pulsations as Probes of
Stellar Physics, ASP Conf. Ser. 259, eds C. Aerts, T. R. Bedding, \& J.
Christensen-Dalsgaard (ASP: San Francisco), 288
\reference Michaud, G., Vauclair, G., \& Vauclair, S. 1983, \apj, 267, 256
\reference Proffitt, C. R., \& Michaud, G. 1991, \apj, 371, 584
\reference Richer, J., Michaud, G., Iglesias, C. A., Rogers, F. J., Turcotte, S., \& LeBlanc, F. 1998, \apj, 492, 833
\reference Richer, J., Michaud, G., \& Turcotte, S., 2000, \apj, 529, 338
\reference Ryabchikova, T. A. 1991, in Evolution of Stars: The
Photospheric Abundances Connection, IAU symp. 145, eds G. Michaud \& A.
Tutukov (Kluwer: Dordrecht), 149
\reference Ryabchikova, T. A., Piskunov, N. E., Kochukov, O., Tsymbal, V., Mittermayer, P., \& Weiss, W. W. 2002, \aap, 384, 545
\reference Vauclair, S., Dolez, N., \& Gough, D. O. 1991, \aap, 252, 618

\section*{Discussion}

\noindent
{\it Ryabchikova:} Do you expect different abundance profiles for
example, for Fe-peak elements, in Ap and Am star of the same effective 
temperature?\\

\noindent
{\it Turcotte:} Abundance profiles are expected to differ in Ap and Am stars of the same temperature
mainly because the atmospheres of Am stars is thought to be mixed, as the hydrogen convection zones in Am stars 
are not suppressed by magnetic fields.
In Ap stars, where the atmosphere is not mixed, stratification occurs.
In the interior, one should expect that diffusion in Ap stars will occur as in Am stars because the magnetic 
field is expected to be too small to have a significant effect.\\

\noindent
{\it Cally:} In magnetic stars with complex magnetic structure in the
atmosphere, we might expect a hierarchy of different closed magnetic
loop structures. Might this lead to horizontal chemical inhomogeneities
at the footprints, since we might expect more settling in high loops
compared to that in low loops?\\

\noindent
{\it Turcotte:}  Wherever such loops exist we would expect the inhomogeneities to reflect the
structure of the magnetic field, whatever the geometry. 
One would need much better modeling than available to make specific predictions.\\

\end{document}